\documentclass[]{interact}

\usepackage{epstopdf}
\usepackage[caption=false]{subfig}

\usepackage[numbers,sort&compress]{natbib}
\bibpunct[, ]{[}{]}{,}{n}{,}{,}
\renewcommand\bibfont{\fontsize{10}{12}\selectfont}

\theoremstyle{plain}

\theoremstyle{definition}

\theoremstyle{remark}

\begin{document}


\title{From the Light Quantum to the Photon\\ \small
The Evolution of a Physical Concept}

\author{
\name{A. Collavini\textsuperscript{a,b,f}, V. Bologna\textsuperscript{c,f}, F. Longo\textsuperscript{c,f}, S. Ansoldi\textsuperscript{c,d,f}, F. Parmigiani\textsuperscript{c,e}\thanks{CONTACT F. Parmigiani. Email: fulvio.parmigiani@elettra.eu} }
\affil{\textsuperscript{a}Department of Mathematics, Computer Science and Physics, University of Udine, Udine, ITA \textsuperscript{b}Department of Philosophy, University of Geneva, Geneva, SWI \textsuperscript{c} Department of Physics, University of Trieste, Trieste, ITA \textsuperscript{d} Institute for Fundamental Physics of the Universe, Trieste, ITA \textsuperscript{e} Elettra–Sincrotrone Trieste, Basovizza, ITA \textsuperscript{f} National Institute of Nuclear Physics, Section of Trieste, ITA}
}

\maketitle

\begin{abstract}
This work examines the physical and conceptual evolution of the light quantum from Planck's blackbody theory to the theoretical and experimental developments that led to the quantization of the electromagnetic field. The present study shows that the decisive transition occurred in Einstein's quantum theory of
radiation (1916--1917). Absorption, stimulated emission, and spontaneous emission were formulated as elementary probabilistic mechanisms whose statistical balance alone reproduces the blackbody spectrum. In particular, spontaneous emission requires the emission of a light quantum, thereby implicitly proving its physical necessity before its theoretical status was clarified.
At the same time, the already existing term \emph{photon} began to acquire a stable usage following Lewis's 1926 proposal and became increasingly associated with Einstein's light quantum.
By the mid-1920s, the central problem had shifted from whether light quanta were physically required to how radiation could be incorporated into the emerging quantum-mechanical formalism. This transition marks a key stage, illustrating how initial debates about the existence of light quanta gave way to their integration into a comprehensive theoretical structure. The resulting asymmetry between the novel quantum description of matter and the still-classical description of radiation, called into question by the phenomenon of spontaneous emission, identifies the physical problem that led to the quantization of the electromagnetic field.
\end{abstract}

\section{Introduction}

In modern physics, few ideas have undergone such a profound transformation as the light quantum.
From Planck's theory of blackbody radiation to the modern photon in quantum electrodynamics, the physical meaning attributed to the quantum of electromagnetic radiation has repeatedly changed as quantum theory itself has developed. Understanding this transformation is important for reconstructing the foundations of modern quantum optics, quantum electrodynamics, and, more generally, quantum field theory. \\
This history has been examined from different perspectives \cite{Hentschel2018,Shore2020,Nautiyal2026}, yet the development of the underlying physics and the stabilization of its terminology followed distinct paths. Although the terms \emph{light quantum} and \emph{photon} are now commonly used interchangeably, with the latter having almost completely replaced the former \cite{Hentschel2018,Shore2020, Kragh2014Photon}, they originated in different scientific contexts and only gradually acquired the meanings they have today \cite{Kragh1999,Hentschel2018}.
Distinguishing the development of the underlying physics from the subsequent stabilization of its terminology is therefore essential for understanding how the contemporary concept of the photon emerged. 
At the same time, some of the deeper physical implications of the theories and experiments that established the physical necessity of the light quantum before field quantization have not always been fully explored or explicitly recognized.\\

In 1900, Planck introduced the quantum hypothesis through a statistical model in which matter was represented by hypothetical resonators (\emph{Resonatoren}), whose allowed energies were restricted to integer multiples of the elementary quantity $h\nu$, where $h$ a new constant and $\nu$ the resonator frequency \cite{Planck1900,Planck1901}.\\ 
Five years later, Einstein extended Planck's elementary quantum to electromagnetic radiation itself, proposing that radiant energy is distributed discontinuously in space in discrete quanta of energy $h\nu$ \cite{Einstein1905}. Albeit useful in accounting for the photoelectric effect and related phenomena, the light-quantum hypothesis remained controversial for many years\footnote{The reservations of several leading physicists are illustrated by the report supporting Einstein's election to the Prussian Academy of Sciences in 1913, where Planck and his colleagues acknowledged that Einstein might have "overshot the mark" in introducing light quanta while emphasizing that genuine scientific progress necessarily requires intellectual boldness. Literally, the report states: "\emph{Daß er in seinen Spekulationen zuweilen über das Ziel hinausgeschossen haben mag, wie z.B. in seiner Hypothese der Lichtquanten, wird man ihm nicht allzu schwer anrechnen dürfen; denn ohne ein gewisses Wagnis läßt sich auch in der exaktesten Naturwissenschaft kein wirklich Neues erreichen}". Authors' translation:"That he may occasionally have overshot the mark in his speculations, as for example in his hypothesis of light quanta, should not be held too much against him; for without a certain willingness to take risks, no genuinely new achievement can be attained, even in the most exact natural science".} \cite{Stachel1989,Pais1982}.

Interestingly, although Planck's constant $h$ and the idea of the light quantum emerged from the same historical context, their subsequent developments followed markedly different trajectories. Agreement between theory and experiment rapidly established the universal character of $h$, whose physical significance continued to deepen with the development of quantum theory \cite{Kuhn1978,Kragh1999}. The physical meaning attributed to the light quantum, by contrast, continued to change while its status as a physical entity remained conceptually unsettled.

The remarkable success of classical electrodynamics in accounting for interference, diffraction, and polarization made the acceptance of localized light quanta particularly problematic. This difficulty was further compounded by the corpuscular interpretation naturally suggested by the light quantum, which attributed particle-like properties to radiation without providing a classical material particle to which such properties could be assigned.
This tension also reveals a deeper asymmetry between the two developments.
Planck's quantization modified the description of the allowed energies of matter at atomic level. In contrast, Einstein's light-quantum hypothesis challenged the unrestricted applicability of the continuous electromagnetic description of radiation \cite{Stachel1998Quantum}.

A major development occurred with Einstein's quantum theory of radiation in 1916--1917 \cite{Einstein1917}. 
Einstein approached the problem of radiative equilibrium by considering how matter with a discrete energy structure could exchange energy with radiation through elementary transitions.
Yet his construction retained an explicit conceptual link with Planck's resonators. In introducing the laws governing these transitions, Einstein stated that his hypotheses were obtained by ``carrying over the known situation for a Planck resonator in classical theory to the as yet unknown one in quantum theory'' \cite{Einstein1917}. The classical resonator thus served as a heuristic guide for formulating the elementary quantum mechanisms of radiative exchange. Einstein introduced three such mechanisms: absorption, stimulated emission, and spontaneous emission. The analogy is particularly revealing for spontaneous emission: just as an oscillating Planck resonator radiates without requiring excitation by an external field, Einstein postulated that a transition from an upper to a lower quantum state could occur with emission of radiation “without excitation by an external cause” \cite{Einstein1917}. This distinctive feature of spontaneous emission gives it a role in Einstein's theory that extends beyond the recovery of radiative equilibrium.\footnote{Einstein explicitly recognized the conceptual difficulty implied by
this description. His theory assigned a probability to the elementary
spontaneous transition but did not determine the time or direction of an
individual emission event, which he described as being left to ``chance''
(\emph{Zufall}) \cite{Einstein1917}. He regarded this indeterminacy as a limitation of the theory, not as an established abandonment of causality.}\\
The present work aims to identify, through the interplay between experimental evidence and theoretical interpretation, how the light quantum evolved from a heuristic hypothesis into a physical entity required by both theory and experiment. We argue that the decisive theoretical transition occurred in Einstein's radiation theory of 1916--1917, when spontaneous emission made the light quantum necessary for radiative equilibrium.\\
A central implication of this study is that the physical necessity of the light quantum emerged before the theoretical framework capable of explaining its origin. The experimental evidence for localized quantum events accumulated during the 1920s reinforced this necessity, while leaving unresolved the fundamental question of how radiation itself could be incorporated into the quantum-mechanical formalism. The emergence of this unresolved physical problem provided the missing link that eventually brought the electromagnetic field itself within the domain of quantum mechanics.
\section{Einstein's Two Radiation Theories: 1905 and 1916--1917}
\label{sec:einstein-radiation}

Einstein's 1905 and 1916--1917 contributions to the theory of radiation
are commonly regarded as successive milestones in the historical development
of the light quantum \cite{Pais1979,Hentschel2018}. They addressed different physical problems and assigned correspondingly different roles to radiation quanta. The distinction between these roles is essential for understanding the change in the physical status of the light quantum that occurred between the two formulations.
Einstein's 1905 paper occupies a unique position in the early development of quantum theory. Unlike Planck, Einstein focused on the broader question of whether several apparently unrelated optical phenomena could be understood more naturally if radiant energy itself possessed a discontinuous structure \cite{Einstein1905,Norton2006Miraculous}.

The opening paragraph of Einstein's 1905 paper reveals the tentative nature of the proposal:

\begin{quote}
\small
\textit{Es scheint mir, daß die mit der schwarzen Strahlung, der Fluoreszenz, der Erzeugung von Kathodenstrahlen durch ultraviolettes Licht und anderen Verwandlungs- oder Erzeugungsvorgängen des Lichtes zusammenhängenden Beobachtungen sich leichter verstehen lassen, wenn man annimmt, daß die Energie des Lichtes im Raume diskontinuierlich verteilt ist\footnote{English translation by the authors:"It seems to me that the observations associated
with blackbody radiation, fluorescence, the production of cathode rays by ultraviolet light, and other related phenomena connected with the emission or transformation of light are more readily understood if one assumes that the
energy of light is discontinuously distributed in space".}.
}
\end{quote}

The wording is itself significant.\\
Building on Planck's radiation law, Einstein showed that, in the Wien regime,
monochromatic radiation behaves thermodynamically as if it consisted of
independent energy quanta of magnitude $E_\gamma=h\nu$, thereby attributing
discreteness to radiation itself\footnote{Einstein formulates the conclusion
of his thermodynamic argument in explicitly cautious terms:
\emph{"Monochromatische Strahlung von geringer Dichte [...] verhält sich
in wärmetheoretischer Beziehung so, wie wenn sie aus voneinander unabhängigen
Energiequanten von der Größe $R\beta\nu/N$ bestünde".}.
In modern notation, $R\beta/N=h$, so that the energy of each quantum is
$h\nu$. A direct translation reads: ``Monochromatic radiation of low density
[...] behaves, from a thermodynamic point of view, as if it consisted of
mutually independent energy quanta of magnitude $h\nu$.'' The expression
\emph{``wie wenn''} (``as if'') is particularly significant, reflecting the
heuristic status that Einstein still assigned to the light-quantum hypothesis
at this stage \cite{Einstein1905,Norton2006Miraculous}.}
The subscript $\gamma$ is introduced to distinguish this quantity
from the quantized energies of Planck's material resonators, for which the
frequency $\nu$ refers to the resonator frequency.

The heuristic character of Einstein's proposal also helps explain its difficult reception. Although the photoelectric effect was consistent with the relation $E_\gamma=h\nu$, it did not by itself establish the light quantum as a necessary constituent of a general theory of radiation, particularly since classical wave theory continued to account successfully for interference, diffraction, and polarization.
As emphasized by Kuhn, early quantum theory was characterized by the coexistence of partially incompatible physical descriptions \cite{Kuhn1978}. Accordingly, the interpretation of the light quantum as a physically real, particle-like entity remained far from generally accepted.
Instead, it could still be regarded as a useful expedient for accounting for specific experimental observations.

From today's perspective, it is important not to identify Einstein's 1905 light quantum retrospectively with the modern photon. The crucial question that remained open was whether this empirical picture could be incorporated into a consistent theory of radiative processes.
An important intermediate development came from Einstein's analysis of
radiation fluctuations in 1909. The resulting fluctuation formula contained
distinct terms associated with the continuous wave properties and the discrete
energy structure of radiation, indicating that both aspects entered its
statistical behavior\footnote{Einstein showed that the mean-square energy
fluctuation of blackbody radiation contains two additive contributions of
different physical character: a term corresponding to the fluctuations
expected from a classical wave field and a term proportional to $h\nu$,
associated with the discrete energy structure of radiation. He interpreted
their simultaneous occurrence as indicating that the wave and corpuscular
descriptions capture distinct aspects of radiation rather than mutually
exclusive alternatives. This result did not amount to a quantization of the
electromagnetic field, but exposed particularly clearly the unresolved
coexistence of its continuous and discrete properties
\cite{Einstein1909Radiation}.}.
In the same year, G. I. Taylor showed that diffraction fringes persist even when the incident light is attenuated to extremely low intensities, approaching conditions in which only very few indivisible energy units could be present in the apparatus at a given time. Although the experiment did not employ a true single-photon source, it provided an early and striking experimental indication that the wave-like propagation of light persists even in the regime of extremely weak radiation \cite{Taylor1909}.
These results further exposed the conceptual difficulty of reconciling the
light quantum with the classical wave description of radiation.\\
A few years later, Einstein returned to the problem from a different perspective. By then, Planck's constant had become one of the central quantities of the emerging quantum theory. Its appearance in the photoelectric effect, the quantum theory of specific heats, Bohr's atomic model, and X-ray spectroscopy had progressively established its universal significance \cite{Bohr1913,Jammer1966,Pais1982,Kragh1999}. 
The existence of elementary quantum phenomena was no longer the central issue. The problem had become understanding the physical role of the light quantum in the elementary interactions between radiation and matter.
Rather than considering an ensemble of hypothetical resonators, Einstein investigated the exchange of energy between radiation and molecules (\emph{Moleküle}) in discrete quantum states under conditions of thermal equilibrium  \footnote{Einstein formulated the argument in terms of a gas molecule
that could occupy different ``quantum theoretical states'' $Z_1,Z_2,\ldots$,
and then considered radiative transitions between a selected pair,
$Z_n$ and $Z_m$ \cite{Einstein1917}. Thus, the construction should not
be understood as assuming a two-level atom in the modern sense.} \cite{Einstein1917}. The focus thus shifted from the statistical properties of quantized material resonators to the elementary processes through which matter and radiation interact. Einstein introduced three radiative processes: absorption, stimulated emission, and spontaneous emission. Each involving a radiative transition between lower to upper energy stationary states (absorption) and vice versa (stimulated emission and spontaneous emission), whose energy difference satisfies $E_2-E_1=h\nu$.
In Einstein's theory, spontaneous emission is a necessary condition for radiative equilibrium. 
Remarkably, in this case, a quantized material system in its excited state must be able to undergo a transition to the lower state even in the absence of an external stimulus, with a probability independent of the radiation density.
The corresponding energy difference,
$E_2-E_1=h\nu=E_\gamma$, is then released into radiation as a light quantum\footnote{In Einstein's 1916--1917 formulation, this distinction is
phenomenological: the probability of spontaneous emission is independent of
the radiation density, whereas those of absorption and stimulated emission
depend on it. No physical mechanism responsible for the spontaneous transition
is specified by the theory. This statement should not be interpreted
retrospectively in terms of vacuum fluctuations or of the interaction with a
quantized electromagnetic field, concepts that emerged only with later
developments of quantum theory.}.

This establishes a logical implication that goes beyond the heuristic use of the light quantum. If spontaneous emission is a necessary elementary mechanism to reproduce the blackbody spectrum, and if every spontaneous transition entails the emission of a light quantum, $E_\gamma$, the light quantum therefore becomes an indispensable physical entity in the description of elementary radiative processes.\\
A separate question concerns the physical origin of spontaneous emission. Einstein's theory characterizes the spontaneous transition phenomenologically through a transition probability, but it does not provide a mechanism determining the occurrence of the individual event.
This problem, already recognized by Einstein - who drew an explicit analogy between the statistical law of spontaneous emission and Rutherford's law of radioactive decay\footnote{Einstein explicitly compared the statistical law governing
spontaneous emission with Rutherford's law of radioactive decay. In both
cases, a constant transition probability per unit time leads to an
exponential survival law,
\[
N(t)=N_0 e^{-\lambda t},
\]
where $\lambda$ denotes the decay or transition rate. The analogy concerns
the statistical structure of the two phenomena, not their physical
mechanisms: the law determines the evolution of an ensemble but does not
specify when an individual radioactive decay or spontaneous-emission event
will occur.} - exposed a fundamental departure from the deterministic description characteristic of classical physics\footnote{Pais has emphasized the significance of this point, identifying Einstein's 1916--1917 radiation theory as the beginning of his concern with the breakdown of classical causality in quantum phenomena \cite{Pais1979}.} \cite{Pais1979}.

Remarkably, despite their different physical approaches, Planck's and Einstein's theories reproduce the same blackbody spectrum. This shows that the Planck distribution is not tied to a unique physical description of the mechanisms underlying radiative equilibrium.
Einstein's radiation theory thus introduced two closely related changes: radiative equilibrium was described through the statistical balance among three elementary probabilistic processes, while the light quantum acquired a necessary physical role within those processes.

\section{From Heuristic Hypothesis to Experimental Necessity}
\label{subsec:instability}

Einstein's radiation theory did not resolve the fundamental tension that had accompanied the light-quantum hypothesis since 1905. Although discrete radiative events had acquired a necessary role within the theory, their apparent incompatibility with the well-established continuous wave properties of light remained unresolved\footnote{The continuous nature of electromagnetic fields in Maxwell's electrodynamics reflects the fact that classical electrodynamics is intrinsically formulated as a continuum theory.} \cite{Kuhn1978}. 
This difficulty helps explain why several leading physicists of the
period---including Planck, Lorentz, Sommerfeld, and later Millikan---remained
skeptical of the physical reality of light quanta despite the steadily growing
experimental evidence \cite{Kragh1999,Millikan1916}.

The intellectual climate of the period is illustrated by Arnold Sommerfeld's
remarks at the 1911 meeting of the Society of German Scientists and Physicians,
where he described the theory of energy quanta as a ``problematic current issue,''
emphasizing that its theoretical foundations were still unsettled
\cite{Sommerfeld1911Quantum}.

A particularly revealing example is provided by Robert Millikan. His photoelectric measurements provided an independent determination of Planck's constant, experimentally confirming, with unprecedented precision, that $h$ also governs the relation between the energy of photoelectrons and the frequency of the incident radiation\footnote{Millikan reported a photoelectric determination
$h=6.57\times10^{-27}\,\mathrm{erg\,s}$ with "a precision of about
0.5 per cent" \cite{Millikan1916}.}. Yet, despite confirming Einstein's photoelectric equation with remarkable precision, Millikan continued to reject the light-quantum hypothesis, describing it in 1916 as ``bold, not to say reckless,'' because a localized electromagnetic corpuscle appeared incompatible with the established theory of optical interference \cite{Millikan1916}.

By the early 1920s, the nature of the debate had changed substantially. The existence of elementary quantum phenomena had become increasingly difficult to deny, but this did not yet imply the acceptance of localized radiation quanta as physically real. What remained unresolved was how the growing body of quantum evidence could be reconciled with the continuous electromagnetic field description.

The most systematic attempt to preserve the continuous character of classical light was the Bohr--Kramers--Slater (BKS) theory \cite{BKS1924,Kragh1999}. Rather than following the physical implications of Einstein's radiation
theory to the conclusion that light quanta are essential to individual
radiative events\footnote{In Einstein's radiation theory, energy and momentum are associated
with each individual elementary radiative event. For a transition satisfying
$E_2-E_1=h\nu$, the exchanged radiant energy is $E_\gamma=h\nu$, while the
corresponding momentum has magnitude $p_\gamma=h\nu/c$. In absorption, the
momentum transferred to the material system is directed along the incident
radiation; in emission, it is directed oppositely. Einstein further concluded
that spontaneous emission occurs in a definite but randomly selected direction
for each individual event, with the emitting system undergoing a recoil of
magnitude $h\nu/c$ in the opposite direction \cite{Einstein1917}.}, 
Bohr, Kramers, and Slater retained a continuous electromagnetic field and
proposed that energy and momentum conservation need hold only statistically
over many elementary events. In this way, the BKS theory sought to preserve the classical wave description of radiation while accommodating the increasing evidence for quantum behavior.

This possibility was decisively tested experimentally. Coincidence measurements by Bothe and Geiger, later confirmed independently by Compton and Simon, showed that energy and momentum are conserved in individual scattering events \cite{BotheGeiger1925,ComptonSimon1925}. These experiments removed the principal experimental support for the BKS theory and established that each radiation--matter interaction constitutes a distinct elementary event.

The significance of these experiments goes beyond the rejection of the BKS theory. The light quantum, introduced in 1905 had now evolved in a firmer physical status, since light quanta and the associated localized quantum events had become indispensable for interpreting the experimental evidence. In this sense, theoretical necessity preceded experimental necessity. By the mid-1920s, a purely continuous description of radiation had become increasingly difficult to sustain.\\
At almost the same time, a complementary development was taking place in the description of matter. De Broglie's matter-wave hypothesis of 1923--1924 provides an instructive parallel to the evolving physical status of the light quantum\footnote{Einstein was not formally involved in the examination of de Broglie's
doctoral thesis. According to de Broglie's later recollection, Langevin sent
Einstein a copy of the thesis and Einstein expressed a strongly favorable
opinion of the work; see \cite{Pais1979}.} \cite{deBroglie1923,deBroglie1924,Jammer1966}. He extended to material particles the association between energy and frequency that had emerged in the theory of radiation, thereby attributing wave properties to entities whose corpuscular character and finite mass were well established. \\
The historical problem of radiation had developed in the opposite direction: the wave character of light was firmly supported by interference and diffraction, while the physical reality of localized light quanta remained controversial. The two developments were not symmetrical. De Broglie's proposal did not by itself resolve the physical status of the light quantum, but it contributed to the broader transformation through which wave and particle descriptions could no longer be regarded as mutually exclusive alternatives. This transformation received striking experimental support with the observation by Davisson and Germer of electron diffraction from a nickel crystal, which experimentally demonstrated the wave behavior of massive particles and confirmed a central prediction of de Broglie's matter-wave hypothesis \cite{DavissonGermer1927}.

During the crucial 1920s, as quantum mechanics was taking shape, reconciling the particle-like properties of light quanta with the wave character of radiation became a central physical problem. The physical necessity of localized quantum events had become increasingly difficult to dispute, but their theoretical interpretation remained incomplete, and the terminology used to describe them had not yet stabilized. Experimental necessity did not by itself produce the modern concept of the photon. Rather, the physical evidence, the emerging quantum theory, and the language used to describe the quantum of radiation were still developing along partially distinct paths. It is within this context that the word \emph{photon} entered physics.

\section{Terminological Disambiguation: From the Light Quantum to the Photon}
\label{sec:terminology}
The term \emph{photon} acquired its lasting place in physics following Gilbert
N. Lewis's 1926 proposal, although the word itself had already appeared earlier
in other, largely unrelated scientific contexts\footnote{The word
\emph{photon} was not a neologism introduced for the first time by Lewis.
Kragh has documented several earlier uses of the term, beginning with
L.~T.~Troland in 1916 and followed by J.~Joly, R.~Wurmser, and F.~Wolfers.
These earlier usages arose in different contexts and were soon forgotten;
see \cite{Kragh2014Photon}.}. 
Lewis's photon had a physical meaning distinct both from Einstein's light quantum
and from the quantum of radiation that would later emerge from the quantization
of the electromagnetic field.

Lewis introduced the term in the context of a theory in which radiation involved
the exchange of hypothetical conserved entities between atoms
\cite{Lewis1926}. His proposal was therefore not simply a new name for
Einstein's \emph{Lichtquant}.

Lewis made this distinction explicit:

\begin{quote}
\small
\textit{``I therefore take the liberty of proposing for this hypothetical new atom, which is not light but plays an essential part in every process of radiation, the name photon.''}
\end{quote}

\begin{flushright}
--- Gilbert N. Lewis (1926) \cite{Lewis1926}
\end{flushright}
Lewis's formulation leaves little ambiguity about this distinction. His photon was explicitly “not light” and was conceived as a conserved entity associated with radiative exchange.
Furthermore, Lewis's construction was not introduced in response to a specific experimental anomaly, nor was it required to account for an established body of experimental evidence. It was essentially a speculative physical model, without direct experimental support, in which the photon was postulated as an indestructible carrier involved in the exchange of radiant energy \footnote{Lewis's description of the photon as a carrier should not be
interpreted as an anticipation of the modern role of the photon as the
mediator of electromagnetic interactions. The resemblance is only superficial:
the two concepts have entirely different theoretical origins and physical
meanings.}.

This historical circumstance reveals an important distinction between the evolution of scientific terminology and the evolution of the physics to which that terminology refers. 
The change in meaning occurred remarkably rapidly. Already in 1927, Arthur Compton used the term \emph{photon} to denote the light quantum involved in X-ray scattering, entirely independently of Lewis's original interpretation. The terminology was sufficiently established that the Fifth Solvay Conference, held in Brussels in October 1927, was subsequently published under the title \emph{Électrons et photons} \cite{Solvay1928}. In his Nobel Lecture delivered in December of the same year, Compton likewise referred to the X-ray quantum as a \emph{photon} and to the corresponding corpuscular description as the ``photon theory'' \cite{Compton1927Nobel}. Thus, within approximately one year of Lewis's proposal, the word had already become detached from the physical model for which it had originally been introduced and was being used as a new name for Einstein's light quantum \cite{Kragh1999,Scully1972}. \\
The persistence of the same word across changing theoretical contexts could suggest a continuity that did not exist at the level of physical interpretation. Planck's energy quantum, Einstein's light quantum, Lewis's photon, and the later quantum of the electromagnetic field did not represent an unchanged physical entity described by different names; they emerged within different experimental and theoretical frameworks and fulfilled different physical functions.
The stabilization of the term did not, however, resolve the physical problem that the light quantum still posed.\\
At the same time, Bose's statistical treatment of blackbody radiation also had formulated the problem in terms of light quanta \cite{Bose1924}, while advances in quantum mechanics were transforming the theoretical description of matter at atomic level.
These developments left a fundamental asymmetry unresolved. Matter could now be described by quantum mechanics models, whereas radiation still occupied an ambiguous theoretical position. Its wave properties were described
by the continuous electromagnetic field, while its exchanges of energy and momentum with matter at a microscopic level required discrete quanta. 

The problem facing the new quantum mechanics was no longer primarily to establish whether light quanta were physically required. Theory and experiment had already made their role increasingly unavoidable. The deeper problem was to incorporate radiation and its interaction with quantum matter
within the same theoretical framework. In particular, a theory was needed in which the continuous electromagnetic field and the discrete exchange of radiation quanta were not introduced as separate elements, but emerged from a common probabilistic formulation.

This unresolved problem became part of the broader theoretical effort to incorporate radiation into the new quantum-mechanical framework. Quantum mechanics could already describe matter and its interaction with an external electromagnetic field \cite{Dirac1926,Slater1927}, but radiation itself remained outside the quantum description. 
A particularly significant example was provided by Wentzel's 1926 quantum-mechanical treatment of the photoelectric effect. By describing the electronic states quantum mechanically while retaining the incident radiation as a classical electromagnetic field, Wentzel showed that the photoelectric effect itself did not require a quantized radiation field. The discreteness of the observed electron transitions could arise from the quantum structure of matter interacting with classical radiation \cite{Wentzel1926Photoelectric}. Viewed retrospectively, his treatment provides an early realization of the semiclassical framework in which quantized matter interacts with classical radiation, i.e., absorption and stimulated emission. In this respect, it also illustrates how far Planck's original idea of restricting quantization to matter could be carried within the new quantum mechanics: absorption, and more generally stimulated radiative transitions, can indeed be described without quantizing the electromagnetic field.
The limitation of this picture becomes apparent with spontaneous emission. Einstein's radiation theory had already shown that this mechanism is required, together with absorption and stimulated emission, to recover the blackbody spectrum, whereas it cannot arise from the Schrödinger dynamics of matter interacting solely with a classical electromagnetic field. The problem was therefore not whether some radiation-matter interactions could be described without light quanta - they could - but whether a theory retaining a classical electromagnetic field could provide a complete account of radiative processes.
Dirac explicitly recognised this limitation, observing that “hardly anything has been done up to the present on quantum electrodynamics\footnote{The expression \emph{quantum electrodynamics} was used
by Dirac in his 1927 paper and is generally attributed to him. The terminology
was rapidly adopted: Jordan and Pauli used \emph{Quantenelektrodynamik}
in 1928, while the subsequent work of Heisenberg and Pauli contributed to
establishing quantum electrodynamics as a distinct theoretical framework
\cite{Dirac1927}.}". The problem was thus to extend the quantum-mechanical formalism to the electromagnetic field itself. Within the broader theoretical effort to achieve this extension, Dirac's 1927 theory of the emission and absorption of radiation represented a decisive step \cite{Dirac1927}.

\section{Conclusions}
The central result of this work is that the physical necessity of the light
quantum emerged before a theoretical framework capable of accounting for its
origin became available. Rather than viewing its historical development as a
linear progression toward the photon or as a mere evolution of terminology,
we have distinguished the different physical meanings acquired by the light
quantum, from Planck's quantum hypothesis to Einstein's theory of radiation
and to the conceptual conditions that ultimately led to the quantization of
the electromagnetic field.

Einstein's radiation theory marked a decisive step in this development.
The emitted light quantum became necessary for radiative equilibrium, while
the experiments of the following decade made discrete quantum events
unavoidable in elementary radiation--matter interactions. By the mid-1920s,
the question was therefore no longer whether light quanta were physically
required, but how such discrete radiative events could arise from a theory
in which matter was described quantum mechanically while radiation remained
a continuous classical field.

Spontaneous emission exposed this incompleteness in its sharpest form.
Einstein's theory required it as an elementary process, yet its origin could
not be accounted for by a classical electromagnetic field interacting with
quantised matter. The light quantum was thus already required by both theory
and experiment before the electromagnetic field itself entered quantum
mechanics. The unresolved problem had shifted from establishing the existence
of light quanta to understanding their theoretical origin, thereby providing
one of the conceptual conditions that led to the quantization of the
electromagnetic field.

The evolution of the concept was accompanied by an equally significant
evolution of its language. The term \emph{photon} did not simply replace
\emph{light quantum}. It originated within a different and short-lived
physical model, survived the disappearance of that model, and was subsequently
transferred to a concept whose physical meaning continued to develop.
Terminology therefore became part of the conceptual history of the light
quantum rather than merely a change in the language used to describe it.

The quantization of the electromagnetic field would open a new stage in this
history, but it lies beyond the scope of the present work. It would provide a
theoretical framework for the origin of radiation quanta without exhausting
the question of their physical meaning. In this sense, Einstein's later
question---\emph{what is a light quantum?}---remained meaningful even after
the theoretical framework in which that question had first arisen had
fundamentally changed.

\section*{Acknowledgments}

The authors thank Giorgio Pastore for his careful reading of the manuscript and for suggesting important references that helped improve the work.

\section*{Declaration of AI use}

During the preparation of this manuscript, the authors used ChatGPT (OpenAI) to assist with language editing, clarity and organization of the text, as well as with some reference checking and literature searches. All references and information obtained with AI assistance were independently verified by the authors against the original sources. All AI-assisted content was critically reviewed and edited by the authors, who take full responsibility for the final content of the manuscript.

\newpage
\bibliographystyle{unsrt}

\end{document}